\documentclass[12pt, centerh1]{article}

\usepackage{moreverb}

\usepackage{amssymb, natbib, bm, caption}
\usepackage{lscape,graphics}
\usepackage{mathtools,colonequals}
\usepackage{caption}
\usepackage{theorem}
\usepackage{placeins}
\usepackage{enumerate}
\usepackage{lineno,hyperref}
\modulolinenumbers[5]

\newcommand{\diag}{\,\mbox{diag}}

\newcommand{\load}{\mathbf\Lambda}
\newcommand{\noisev}{\mathbf\Psi}
\newcommand{\Beta}{\mbox{\boldmath$\beta$}}

\newcommand{\vecx}{\mathbf{x}}
\newcommand{\vece}{\mathbf{e}}
\newcommand{\vecE}{\mathbf{E}}
\newcommand{\vecX}{\mathbf{X}}
\newcommand{\vecU}{\mathbf{U}}
\newcommand{\vecR}{\mathbf{R}}

\newcommand{\vecV}{\mathbf{V}}
\newcommand{\vecw}{\mathbf{w}}
\newcommand{\vecz}{\mathbf{z}}
\newcommand{\vecu}{\mathbf{u}}
\newcommand{\vecv}{\mathbf{v}}

\newcommand{\vecmu}{\boldsymbol\mu}

\newcommand{\vecLambda}{\mathbf\Lambda}
\newcommand{\varthet}{\boldsymbol\vartheta}
\newcommand{\vecSigma}{\mathbf\Sigma}
\newcommand{\vecXi}{\mathbf\Xi}
\newcommand{\vecPsi}{\mathbf\Psi}
\newcommand{\vecBeta}{\mathbf\Beta}
\newcommand{\vecTheta}{\mathbf\Theta}
\newcommand{\vecDelta}{\mathbf\Delta}
\newcommand{\vectheta}{\boldsymbol\theta}
\newcommand{\vecepsilon}{\boldsymbol\epsilon}
\newcommand{\vecdelta}{\boldsymbol\delta}

\newcommand{\vecvartheta}{\boldsymbol\vartheta}

\newcommand{\matsig}{\mathbf{\Sigma}}

  \title{A Mixture of SDB Skew-$t$ Factor Analyzers} 
  
  \author{Paula M.\ Murray$^*$, Ryan P.\ Browne$^{**}$ and Paul D.\ McNicholas$^*$}
\date{\small $^{*}$Dept.\ of Mathematics \& Statistics, McMaster University, Hamilton, Ontario, Canada.\\
$^{**}$Dept.\ of Statistics and Actuarial Sciences, University of Waterloo, Ontario, Canada.}

\begin{document}
\maketitle
\begin{abstract}
\noindent Mixtures of skew-$t$ distributions offer a flexible choice for model-based clustering.  A mixture model of this sort can be implemented using a variety of formulations of the skew-$t$ distribution.  Herein we develop a mixture of skew-$t$ factor analyzers model for clustering of high-dimensional data using a flexible formulation of the skew-$t$ distribution. Methodological details of our approach, which represents an extension of the mixture of factor analyzers model to a flexible skew-$t$ distribution, are outlined and details of parameter estimation are provided. Clustering results are illustrated and compared to an alternative formulation of the mixture of skew-$t$ factor analyzers model as well as the mixture of factor analyzers model.\\[-6pt]

\end{abstract}

\section{Introduction}
Mixture models have become an increasingly popular tool for clustering since they were used by \cite{wolfe65}, and the term model-based clustering is commonly used to describe the application of mixture models for clustering. \cite{fraley02a}, \cite{bouveyron14}, and \cite{mcnicholas16b} provide reviews of work in model-based clustering and \cite{mcnicholas16a} deals with the subject in a monograph. Although much of the work on model-based clustering has been based on Gaussian mixtures, recent years have seen extensive work on non-Gaussian mixture model-based approaches. In fact, there has been a veritable explosion of such work over the past few years. Much of this work has focused on non-elliptical mixtures, including mixtures of skew-normal distributions \citep*[e.g.,][]{lin07,lin16}, mixtures of skew-$t$ distributions \citep[e.g.,][]{lin10,vrbik12,vrbik14}, mixtures of skew-$t$-normal distributions \citep{lin14}, mixtures of normal-inverse-Gaussian distributions \citep{karlis09, subedi14,ohagan16}, mixtures of variance-gamma distributions \citep{smcnicholas17}, mixtures of shifted asymmetric Laplace distributions \citep{franczak13}, mixtures of multiple scaled distributions \citep{tortora14,wraith15},  
and mixtures of generalized hyperbolic distributions \citep{browne15,morris15}. 

There are several forms of the skew-normal and skew-$t$ distributions, some of which are discussed by \cite{lee14}. To date, three forms have been studied for model-based clustering. The first, which \cite{lee14} refer to as ``restricted", is that developed by \cite{azzalini96}, and the second, which \cite{lee14} refer to as ``unrestricted", is that of \cite{sahu03}. For the reasons outlined by \cite{azzalini14}, we do not use the names ``restricted" and ``unrestricted" herein; however, we note that an alternative perspective is presented by \cite{mclachlan16}. Hereafter, we will refer to the form of \cite{azzalini96} as classical, and to that of \cite{sahu03} simply as SDB, from the authors' initials. The third form that has been studied for model-based clustering is a special and limiting case of the generalized hyperbolic distribution \citep[see][]{murray14b,murray14a}. Note that the classical and SDB formulations have corresponding skew-normal formulations \citep{azzalini14b}; however, the generalized hyperbolic formulation does not.

\cite{montanari10a} develop a heteroscedastic factor mixture analysis model and \cite{montanari10} use a skew-normal factor analysis model. \cite{murray14b} use the generalized hyperbolic formulation of the  skew-$t$ distribution to develop a mixture of skew-$t$ factor analyzers model. \cite{lin15} use the classical formulation of the skew-$t$ distribution to develop a skew-$t$ factor analysis model. \cite{lin16} use the classical formulation of the skew-normal distribution to develop a skew-normal factor analyzers model; mixtures thereof are used for clustering. Of course, development of a mixture of skew-$t$ factor analyzers would follow similarly via the classical formulation. However, the SDB formulation has yet to be used in the development of a (mixture of) factor analyzers model; herein, we extend the mixture of factor analyzers (MFA) model to the SDB skew-$t$ distribution and investigate whether it is worthy of addition to the model-based clustering ``toolbox".

The SDB skew-$t$ distribution was used by \cite{lin10} in a mixture modelling framework, exploiting a variant of the expectation-maximization (EM) algorithm \citep{dempster77} for parameter estimation.  However, this model requires a Monte Carlo estimation step which, as noted by \cite{lee14}, is computationally expensive.  In an effort to improve computational efficiency of the integral estimation in the E-step for the SDB skew-$t$ mixture model, \cite{lee14} write the integrals in the form of a truncated multivariate-$t$ distribution.  This is subsequently written as a non-truncated $t$-distribution and pre-exisiting statistical packages can be used for computation.  Herein we extend the MFA model \citep{ghahramani97,mclachlan00a} using the SDB skew-$t$ distribution. 

Besides skew-normal and skew-$t$ distributions, the MFA model has been extended to other mixture models that parameterize component skewness. Examples include mixtures of generalized hyperbolic factor analyzers \citep{tortora15} and mixtures of variance-gamma factor analyzers \citep{smcnicholas17}.

\section{Background}\label{sec:back}
\subsection{Mixtures of SDB Skew-$t$ Distributions} \label{sec:MST}

The density of a finite mixture model is given by
\begin{equation}
f(\vecx\mid\vectheta_g)=\sum_{g=1}^{G}\pi_g f_g(\vecx\mid\vectheta_g),
\label{eq:generalmixture1}
\end{equation}
where $\pi_g>0$ is a mixing proportion such that $\sum_{g=1}^{G}\pi_g=1$, and $f_g(\vecx\mid\vectheta_g)$ is the $g$th component density with parameters $\vectheta_g$. The density of the SDB multivariate skew-$t$ distribution is given by
\begin{equation}\begin{split}
&b(\vecx\mid\vecmu_g,\vecSigma_g, \nu_g, \vecDelta_g)=2^pt_p(\vecx;\vecmu_g,\vecSigma_g,\nu_g)\\&\times T_p\bigg(\vecDelta_g\vecSigma_g^{-1}(\vecx-\vecmu_g)\sqrt{\left(\frac{\nu_g+p}{\nu_g+d(\vecx~|~\vecmu_g,\vecSigma_g)}\right)}; \mathbf{0}, \mathbf{I}_p-\vecDelta_g\vecSigma_g^{-1}\vecDelta_g, \nu_g+p\bigg),
\label{eq:skewtdens1}
\end{split}\end{equation}
where $\vecDelta_g=\text{diag}(\vecdelta_g)$ is the skewness, 
$d(\vecx~|~\vecmu_g,\vecSigma_g)=(\vecx-\vecmu_g)'\vecSigma_g^{-1}(\vecx-\vecmu_g)$, 
$t_p(\cdot)$ is the density of a $p$-dimensional $t$-distributed random variable, and $T_p(\cdot)$ is the cumulative distribution function. Using $b(\vecx\mid\vecmu_g,\vecSigma_g, \nu_g, \vecDelta_g)$ in \eqref{eq:skewtdens1} as the component density in \eqref{eq:generalmixture1}, i.e., setting $f_g(\vecx\mid\vectheta_g)=b(\vecx\mid\vecmu_g,\vecSigma_g, \nu_g, \vecDelta_g)$ in \eqref{eq:generalmixture1}, leads to a mixture of SDB distributions. A mixture model of this form has previously been used for clustering by \cite{lin10} and \cite{lee14}.  

\subsection{Mixtures of Factor Analyzers}

The factor analysis model \citep{spearman04} assumes that the variation in $p$ observed variables can be explained by $q$ unobserved or latent variables where $q< p$. Consider independent $p$-dimensional random variables $\vecX_1,\ldots,\vecX_n$. We may write the factor analysis model as $$\vecX_i=\vecmu+\vecLambda\vecU_i+\vecepsilon_i,$$ where $\vecmu$ is the mean, $\vecLambda$ is a $p\times q$ matrix of factor loadings, $\vecU_i$ is a $q$-dimensional vector of latent factors such that $\vecU_i\sim\mathcal{N}(\mathbf{0},\mathbf{I}_q)$ independently, and $\vecepsilon_i\sim\mathcal{N}(\mathbf{0},\vecPsi)$ independently and independent of $\vecU_i$, where $\vecPsi$ is a $p\times p$ diagonal matrix with positive entries.  It follows that the marginal distribution of $\vecX_i$ is multivariate Gaussian with mean $\vecmu$ and covariance matrix $\vecLambda\vecLambda'+\vecPsi$. 

The MFA model \citep{ghahramani97, mclachlan00a} has density of the form 
\begin{equation}
f(\vecx\mid\vecmu_g,\vecLambda_g,\vecPsi_g)=\sum_{g=1}^{G}\pi_g \phi(\vecx\mid\vecmu_g,\vecLambda_g\vecLambda_g'+\vecPsi_g),
\label{eq:mfa}
\end{equation}
where $\phi(\vecx\mid\vecmu_g,\vecLambda_g\vecLambda_g'+\vecPsi_g)$ is the density of a multivariate Gaussian random variable with mean $\vecmu_g$ and covariance matrix $\matsig_g$.
The MFA model is effective for modelling high-dimensional data and has been extended in various ways 
\citep[e.g.,][]{mclachlan07,mcnicholas08,mcnicholas10d,baek10,andrews11a,andrews11b,steane11,baek11,murray14b,murray14a}.

\subsection{The Expectation-Maximization Algorithm}

The expectation-maximization (EM) algorithm \citep{dempster77} is an iterative algorithm for finding maximum likelihood estimates when data are incomplete or are treated as such. The EM algorithm has been widely used for parameter estimation in mixture model based-clustering. The EM algorithm alternates between two steps, an expectation (E) step and a maximization (M) step. In the E-step, the expected value of the complete-data log-likelihood is computed, i.e., updated, conditional on the current parameter estimates. In the M-step, this expected value is maximized with respect to the model parameters, i.e., the parameter estimates are updated. \cite{mclachlan08} give a detailed review of EM algorithms and their application to mixture models.

To formulate clustering problems in the complete-data framework, we introduce indicator variables $Z_{ig}$, where $z_{ig}=1$ if $\vecx_i$ is in component $g$ and $z_{ig}=0$ otherwise. The complete-data are then given by the observed $\vecx_1,\ldots,\vecx_n$ together with the missing $\vecz_1,\ldots,\vecz_n$, where $\vecz_i=(z_{i1},\ldots,z_{iG})$ for $i=1,\ldots,n$. For the MSDBFA models, there is another source of missing data, i.e., the latent variables (see Section~\ref{sec:para}). These latent variables join the observed data and the missing labels to form the complete-data for the MSDBFA models. We fit our MSDBFA models using an alternating expectation-conditional maximization (AECM) algorithm \citep{meng97}, which allows different complete-data at each conditional maximization (CM) step. Details are given in Section~\ref{sec:para}. 

\subsection{Parsimonious Mixtures of Skew-$t$ Factor Analyzers}

\cite{murray14b} introduced a family of mixtures of skew-$t$ factor analyzers for clustering high-dimensional data. They use a formulation of the skew-$t$ distribution that arises as a special and limiting case of the generalized hyperbolic distribution \citep[see][]{browne15}, and they use constraints on the components covariance matrices analogous to those used by \cite{mcnicholas08} in the Gaussian framework. The resulting family of models is referred to as the parsimonious mixtures of skew-$t$ factor analyzers (PMSTFA) family. 
The AECM algorithm 
is used for model fitting and extensive details are given by \cite{murray14b}. Note that to make the analyses in Section~\ref{sec:illustrations} direct comparisons, we use only the PMSTFA model with unconstrained scale matrix, i.e., $\matsig_g=\load_g\load_g'+\noisev_g$, and this model is referred to herein as the MSTFA model. Of course, the MSFTA model is a skew-$t$ analogue of the MFA model and uses the generalized hyperbolic formulation of the skew-$t$ distribution. The mixture of skew-$t$ distributions introduced in Section~\ref{sec:meth} uses the SDB formulation of the skew-$t$ distribution.

\section{Methodology}\label{sec:meth}

\subsection{The MSDBFA Model}\label{sec:md}

We can write a random variable $\vecX$ arising from the SDB skew-$t$ distribution, see \eqref{eq:skewtdens1}, as 
$$\vecX=\vecmu+\vecDelta|\vecV^*|+\vecR^*,$$
where $\vecDelta=\text{diag}(\vecdelta)$ is a skewness parameter and 
\begin{equation}
\begin{bmatrix}
\vecV^* \\
\vecR^* \\
\end{bmatrix}
\sim t_{2p}\left(
\begin{bmatrix}
\mathbf{0} \\
\mathbf{0} \\
\end{bmatrix},
\begin{bmatrix}
\mathbf{I}_p & \mathbf{0}\\
\mathbf{0} & \vecXi\\
\end{bmatrix}, \nu \right)
\end{equation}
with $\vecXi=\vecSigma-\vecDelta^2$. It follows that  $\vecX=\vecmu+(\vecR\mid\vecV> \mathbf{0})$, with
\begin{equation}
\begin{bmatrix}
\vecV \\
\vecR \\
\end{bmatrix}
\sim t_{2p}\left(
\begin{bmatrix}
\mathbf{0} \\
\mathbf{0} \\
\end{bmatrix},
\begin{bmatrix}
\mathbf{I}_p & \vecDelta \\
\vecDelta & \vecSigma \\
\end{bmatrix}, \nu \right).
\end{equation}
By introducing a latent variable $W\sim \text{gamma}(\nu/2,\nu/2)$, it follows that
$\mathbf{V}\mid w\sim\text{HN}_p(({1}/{w})\mathbf{I}_p),$
where $\text{HN}_p(\cdot)$ denotes the half-normal distribution, and
$$\vecX\mid\vecv,w \sim \mathcal{N}\bigg(\vecmu+\vecDelta\vecv,\frac{1}{w}\vecSigma\bigg).$$
Thus, we can write
\begin{equation}
\vecX
= \vecmu+\vecDelta|\vecV|+\frac{1}{W}\vecR,
\label{eq:skewFA}
\end{equation}
where $\vecR\sim\mathcal{N}(\mathbf{0},\vecSigma)$. Recall that the factor analysis model is written as
\begin{equation}\label{eqn:v}
\mathbf{R}=\vecLambda\vecU+\vecepsilon,
\end{equation}
where $\vecU\sim\mathcal{N}(\mathbf{0},\mathbf{I}_q)$ and $\vecepsilon\sim\mathcal{N}(\mathbf{0},\vecPsi)$.  Substituting \eqref{eqn:v} into \eqref{eq:skewFA} gives
$$\vecX = \vecmu + \vecDelta|\vecV|+\frac{1}{W}(\vecLambda\vecU+\vecepsilon).$$
It follows that $\vecX\mid\vecv,w\sim \mathcal{N}(\vecmu+\vecDelta|\vecv|, ({1}/{w})(\vecLambda\vecLambda'+\vecPsi))$. 
Therefore, the density of our mixture of SDB skew-$t$ factor analyzers (MSDBFA) model can be written
\begin{equation*}\begin{split}
f(\vecx\mid\varthet)&=\sum^{G}_{g=1}\pi_gb(\vecx\mid\vecmu_g,\vecLambda_g\vecLambda_g'+\vecPsi_g, \nu_g, \vecDelta_g)\\
&=\sum^{G}_{g=1}\pi_g\gamma(w_{ig}\mid\nu_g/2,\nu_g/2)h(\vecv_{ig}\mid({1}/{w_{ig}})\mathbf{I}_p)\phi(\vecu_{ig}\mid\mathbf{0},\mathbf{I}_q)\\&\qquad\qquad\qquad\qquad\times\phi(\vecx\mid\vecmu_g+\vecDelta_g\vecv_{ig},({1}/{w_{ig}})(\vecLambda_g\vecLambda_g'+\vecPsi_g)),
\end{split}\end{equation*}
where $\gamma(\cdot)$ denotes the density of a gamma distribution, $\phi(\cdot)$ denotes the density of a Gaussian distribution, $h(\cdot)$ denotes the density of a half-normal distribution, and $\varthet$ denotes all model parameters.

\subsection{Parameter Estimation}\label{sec:para}

At each iteration of the AECM algorithm, the expected-value of the complete-data log-likelihood is computed. The complete-data log-likelihood for the MSDBFA model is 
\begin{multline}
l_c(\vecvartheta\mid\vecx,\vecw,\vecv, \vecz)=\sum^{n}_{i=1}\sum^{G}_{g=1}z_{ig}\log\big[\pi_g\gamma(w_{ig}\mid\nu_g/2,\nu_g/2)h(\vecv_{ig}\mid({1}/{w_{ig}})\mathbf{I}_p)\\
\times\phi(\vecu_{ig}\mid\mathbf{0},\mathbf{I}_q)\phi(\vecx\mid\vecmu_g+\vecDelta_g\vecv_{ig},({1}/{w_{ig}})(\vecLambda_g\vecLambda_g'+\vecPsi_g))\big],
\end{multline}
The E-steps of our AECM algorithm require the following expected values: 
\begin{equation*}\begin{split}
&\mathbb{E}[Z_{ig}|\vecx_i] = \frac{\pi_g b(\vecx_i\mid\vecmu_g,\vecLambda_g\vecLambda_g'+\vecPsi_g, \nu_g, \vecDelta_g)}{\sum^{G}_{h=1}\pi_hb(\vecx_i\mid\vecmu_h,\vecLambda_h\vecLambda_h'+\vecPsi_h, \nu_h, \vecDelta_h)}\equalscolon \hat{z}_{ig},\\
&\mathbb{E}[W_{ig}\mid\vecx_i,z_{ig}=1]\equalscolon e_{1,ig},\qquad
\mathbb{E}[W_{ig}\vecV_{ig}\mid\vecx_i,z_{ig}=1]\equalscolon \vece_{2,ig},\\
&\mathbb{E}[W_{ig}\vecV_{ig}\vecV_{ig}'\mid\vecx_i,z_{ig}=1]\equalscolon \vecE_{3,ig},\quad
\mathbb{E}[\log W_{ig}\mid\vecx_i,z_{ig}=1]\equalscolon e_{4,ig}.
\end{split}\end{equation*}
We employ the integral approximation method of \cite{lee14}, which we believe offers advantages in terms of accuracy and computation time in evaluating these intractable expectations when compared to the Monte Carlo EM algorithm of \cite{lin10}.

At the first stage of our AECM algorithm, our incomplete-data include the labels $z_{ig}$, the latent variables $w_{ig}$, and the latent $\mathbf{v}_{ig}$. The location $\vecmu_g$ and the skewness $\vecDelta_g$ are updated via 
\begin{equation*}\begin{split}
&\vecmu_g=\frac{\sum^{n}_{i=1}\left[e_{i,ig}\vecx_i-\vecDelta_g\vece_{2,ig}\right]}{\sum_{i=1}^{n}e_{i,ig}},\\
&\vecDelta_g=\left[\left(\vecLambda_g\vecLambda_g+\vecPsi\right)^{-1}\odot\sum^{n}_{i=1}\vecE_{3,ig}\right]^{-1}\text{diag}\left\{\left(\vecLambda_g\vecLambda_g+\vecPsi\right)^{-1}\sum^{n}_{i=1}\left(\vecx_i-\vecmu_g\right)\vece_{2,ig}'\right\},\\
\end{split}\end{equation*}
respectively, where $\odot$ is the Hadamard product, and the equation 
\begin{equation*}\label{eqn:nuup}
\log\left(\nu_g/2\right)-\psi\left(\nu_g/2\right)-\frac{1}{n}\sum^{n}_{i=1}\left(e_{1,ig}-e_{4,ig}\right)+1=0
\end{equation*}
is solved numerically to obtain the update for $\nu_g$.
The ``sample covariance'' matrix $\mathbf{S}_g$ is updated by
\begin{equation*}\begin{split}
\mathbf{S}_g=\frac{1}{n}\sum^{N}_{i=1}&\left[\vecDelta_g\vecE_{3,ig}\vecDelta_g'-\left(\vecx_i-\vecmu_g\right)\vece_{2,ig}'\vecDelta_g+\left(\vecx_i-\vecmu_g\right)\left(\vecx_i-\vecmu_g\right)'e_{1,ig}-\vecDelta_g\vece_{2,ig}\left(\vecx_i-\vecmu_g\right)'\right].
\end{split}\end{equation*}

At the second stage of our AECM algorithm, the incomplete-data include the labels $z_{ig}$, the latent $w_{ig}$, the latent $\mathbf{v}_{ig}$, and the latent factors $\vecu_{ig}$. In this step, the factor loading matrix $\vecLambda_g$ and the error variance matrix $\vecPsi_g$ are updated via 
$${\vecLambda}_g=\mathbf{S}_g{\Beta}_g'{\vecTheta}_g^{-1} \qquad\text{ and }\qquad {\vecPsi}_g=\diag\{\mathbf{S}_g-{\vecLambda}_g{\Beta}_g\mathbf{S}_g\},$$
respectively, where $\vecBeta_g={\vecLambda}_g'({\vecLambda}_g{\vecLambda}_g'+{\vecPsi}_g)^{-1}$ and $\vecTheta_g=\mathbf{I}_p-{\vecBeta}_g{\vecLambda}_g+{\vecBeta}_g\mathbf{S}_g{\Beta}_g'$. Unsurprisingly, these updates are analogous to those for the MFA model; see \cite{mcnicholas08}, who use similar notation.

\subsection{Initialization and Convergence}

In the illustrations herein (Section~\ref{sec:illustrations}), $k$-means starting values are used for the MSDBFA model as well as all other approaches. Convergence of our EM algorithm is determined using a criterion based on the Aitken acceleration \citep{aitken26}. The Aitken acceleration at iteration $k$ is
\begin{equation}\label{eqn:aa}
a^{(k)} = \frac{l^{(k+1)}-l^{(k)}}{l^{(k)}-l^{(k-1)}},
\end{equation}
where $l^{(k)}$ is the (observed) log-likelihood at iteration $k$. The quantity in \eqref{eqn:aa} can be used to derive an asymptotic estimate (i.e., an estimate of the value after very many iterations) of the log-likelihood at iteration $k+1$: $$l_{\infty}^{(k+1)} = l^{(k)} + \frac{1}{1-a^{(k)}}(l^{(k+1)}-l^{(k)})$$ \citep[see][]{bohning94, lindsay95}. Following \cite{mcnicholas10a}, we stop our EM algorithms when 
\begin{equation}\label{eqn:stopping}
l_{\infty}^{(k+1)}-l^{(k)} < \epsilon,
\end{equation} 
provided this difference is positive. In the analyses in Section~\ref{sec:illustrations}, we use the stopping rule \eqref{eqn:stopping} with $\epsilon=0.01$.

\subsection{Dimension Reduction}

\begin{figure}[ht]
\centering\includegraphics[width=0.65\textwidth,angle=270]{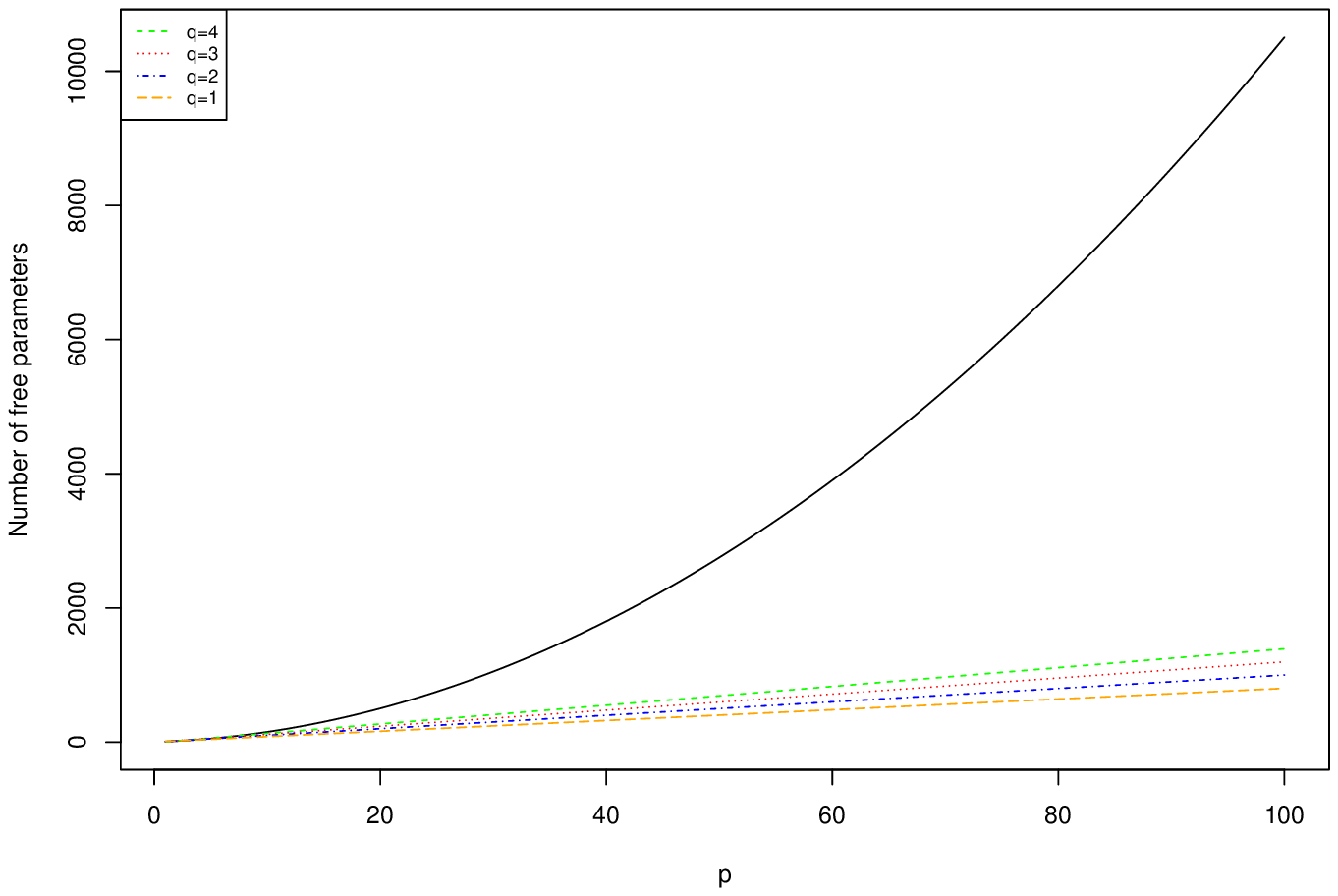}
\caption{A curve depicting the number of free parameters against the number of variables $p$ for the two-component mixture of SDB skew-$t$ distributions, with lines depicting the number of free parameters for the two-component MSDBFA model with $q=1,\ldots,4$.}\label{fig:paras}
\end{figure}

The number of free parameters in the mixture of SBD skew-$t$ distributions is given by $(G-1)+G(2p+1)+Gp(p+1)/2$. In all, $Gp(p+1)/2$ of these free parameters come from the component scale matrices $\matsig_1,\ldots,\matsig_G$. In terms of the number of free parameters, the effect of the MSDBFA model is to reduce the $Gp(p+1)/2$ free parameters in the component scale matrices to  $G[p(q+1)-q(q-1)/2]$. In turn, the impact on the overall number of free parameters is that it reduces from quadratic to linear in $p$ (Figure~\ref{fig:paras}). Note that Figure~\ref{fig:paras} is based on a two-component mixture of SDB skew-$t$ distributions and two-component MSDBFA models; however, changing the number of components only has the effect of changing the values on the $y$-axis.

Finally, it is important to note that the MSDBFA model is not just useful in situations where $p$ is considered large. The MSDBFA model can be useful even when $p$ is not large because
the presence of variables that are not helpful in discriminating groups --- sometimes called discriminative variables --- can have a negative effect on clustering, or classification, performance. This topic is discussed in detail by \citet[][Sec.~4.1]{mcnicholas16a}, who points out that approaches like the MSDBFA model are implicit dimension reduction in contrast to variable selection methods, which are explicit.

\section{Illustrations}\label{sec:illustrations}

\subsection{Model Selection and Performance Assessment}

The Bayesian Information Criterion \citep[BIC;][]{schwarz78} can be used to select the best model in terms of the number of groups $G$ and the number of latent factors $q$. The adjusted Rand index \citep[ARI;][]{hubert85} is used to assess clustering performance relative to the true labels. The ARI is a measure of class agreement between the true class labels and the estimated group memberships. The ARI is 1 for perfect classification, has expected value 0 under random classification, and values less than 0 indicate classification results that are worse than we would expect under random classification.

\subsection{Seeds Data}

\begin{figure}[ht]
\centering\includegraphics[width=0.825\textwidth,angle=270]{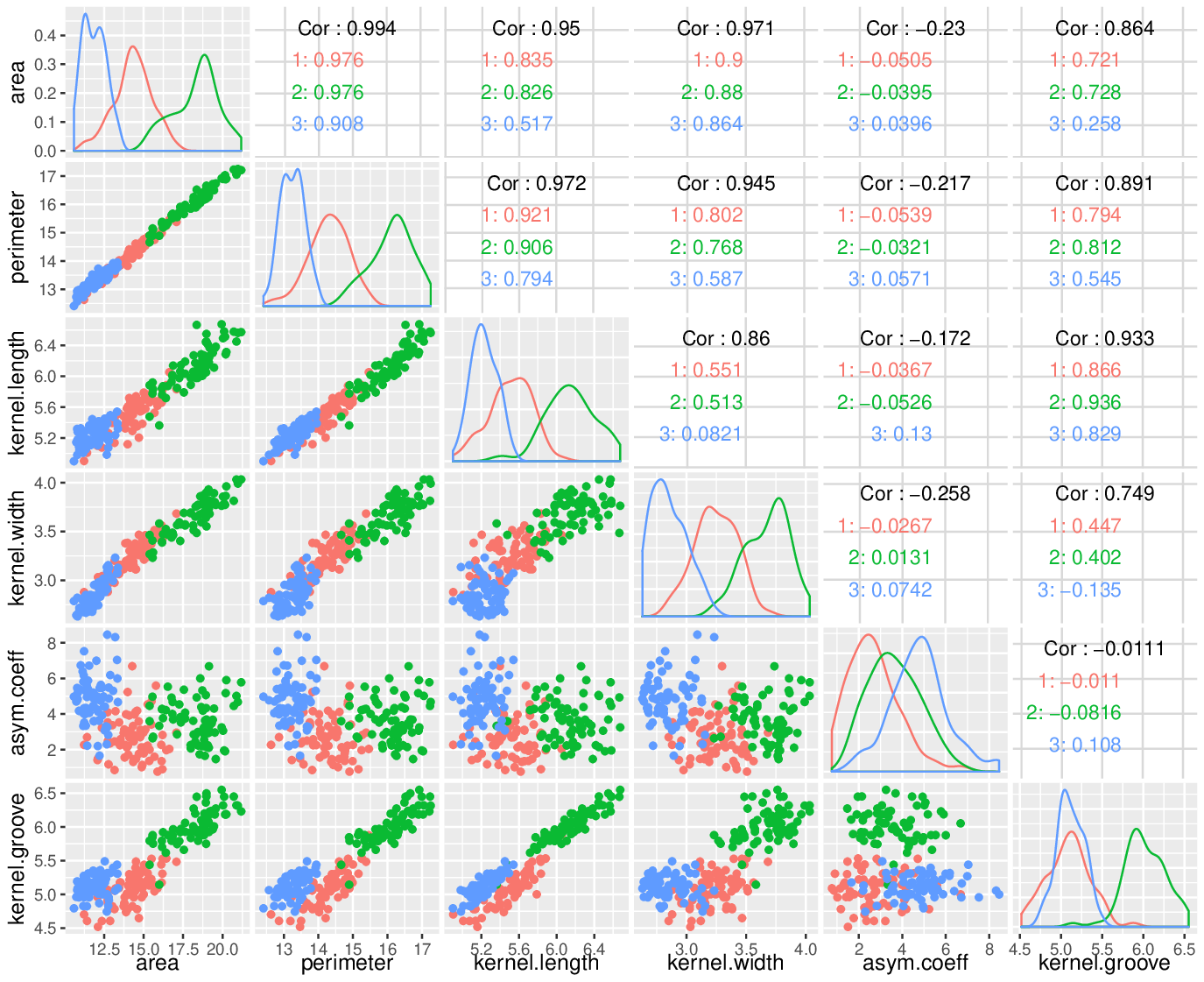}
\caption{Pairs plot for the {\tt seeds} data, with the variable compactness excluded, where colour denotes wheat variety.}\label{fig:seeds}
\end{figure}

We also consider the seeds data set \citep{charytanowicz10} available via the UCI Machine Learning Repository. The data set contains measurements on kernels from three varieties of wheat: Kama, Rosa, and Canadian. In all, there are 210 seeds. 
There are seven measurements but one of them (compactness) is just a function of two others (area and perimeter) via $C = 4\pi A/P^2$. Therefore, the compactness variable is removed. Looking at a pairs plot of the data (Figure~\ref{fig:seeds}), it is clear that area and perimeter are very highly correlated and so area is removed; accordingly, five measurements are considered (perimeter, length, width, asymmetry coefficient, and length of kernel groove). From Figure~\ref{fig:seeds}, it is clear that classes are asymmetric for some of the variables, e.g., kernel width. 

The MFA model is fitted using the {\tt pgmm} package \citep{mcnicholas15} for {\sf R} \citep{R16}. The MSDBFA and MSTFA models are fitting using code written in {\sf R}. Note that when fitting a factor analysis model, the number of model parameters in $\vecSigma$ is only reduced when the relationship 
\begin{equation}
(p-q)^2>(p+q)
\label{eq:factors}
\end{equation}
is satisfied \citep[see][]{lawley71}.
To meet this requirement, the MSDBFA, MSTFA, and MFA models were fit for $G=3$ and $q=1,2$.  Table~\ref{tab:seeds} reports the results from the best models as chosen by the BIC. The MSDBFA model ($\text{ARI}=0.87$) obtains slightly superior clustering results compared to the MFA model ($\text{ARI}=0.84$) and performs better than the MSTFA model ($\text{ARI}=0.53$). Note that the greater BIC value for the MSTFA models illustrates that the model with the best BIC does not necessarily give the best clustering performance.
\begin{table}[!h]
\caption{Clustering results --- ARI and misclassification rate (MR) --- for the best three-component MSDBFA, MSTFA, and MFA models fit to the seeds data.}
\centering
{\begin{tabular*}{\textwidth}{@{\extracolsep{\fill}}lllrrr}
\hline
 & $G$ &$q$&BIC &ARI& MR\\
\hline
MSDBFA&3&1&$-1348.96$&0.87 & 0.043\\
MSTFA& 3& 1& $-1268.72$ &0.53& 0.267\\
MFA& 3& 1& $-1417.25$ &0.84&  0.057\\
\hline
\end{tabular*}}
\label{tab:seeds}
\end{table}

\subsection{Italian Wine Data}

\citet{forina86} recorded chemical and physical properties of three types of wine (Barolo, Grignolino, Barbera) from the Piedmont region of Italy. The {\tt gclus} package \citep{hurley04} contains 13 variables from the original study (Table~\ref{tab:winevars}). 
\begin{table}[!ht]
\caption{Thirteen chemical and physical properties of
Italian wines available in {\tt gclus}.}\label{tab:winevars}
\begin{tabular*}{1.00\textwidth}{@{\extracolsep{\fill}}lll}\hline
{Alcohol} & {Malic acid} & Hue\\
{Magnesium} & {Ash} & {Alcalinity of ash} \\
{Proline} & OD280/OD315 of diluted wines & {Total phenols}\\
{Nonflavonoid phenols} & {Flavonoids} & {Proanthocyanins}\\
Color Intensity &  & \\
\hline
\end{tabular*}
\end{table}

The MSDBFA and MFA models were fitted to the wine data for $G=1,\ldots,4$ and $q=1,2,3$. Table~\ref{tab:wine} reports the results from the best models as chosen by the BIC. The clustering performance of the MSDBFA model ($\text{ARI}=0.98$) is better than for the MFA model ($\text{ARI}=0.91$) and the MSTFA model ($\text{ARI}=0.74$); in fact, the MSDBFA model misclassifies only one observation (Table~\ref{tab:wineclass}).
\begin{table}[!h]
\caption{Clustering results --- ARI and misclassification rate (MR) --- for the best MSDBFA, MSTFA, and MFA models fit to the wine data.}
\centering
\begin{tabular*}{\textwidth}{@{\extracolsep{\fill}}lcccrr} \hline
 & $G$&$q$ & BIC & ARI&MR \\ \hline
MSDBFA &3& 1 & $-4908.28$ & $0.98$ & 0.006\\
MSTFA&3& 1 &$-5352.83$& $0.74$& 0.090\\
MFA &3& 1 & $-5345.56$ & $0.93$ & 0.023\\ \hline
\end{tabular*}
\label{tab:wine}
\end{table}%
\begin{table}[ht]
\caption{Predicted classifications (A, B, C) cross-tabulated against true classes for the best MSDBFA model fitted to the wine data, as selected by the BIC.}\label{tab:wineclass}
\centering
\begin{tabular*}{\textwidth}{@{\extracolsep{\fill}}lrrr}
  \hline
 & A & B & C \\ 
  \hline
Barolo &   50 &   0 &  0 \\ 
Grignolino &   1 &  70 &   0 \\ 
Barbera &  0 &   0 &   48 \\ 
   \hline
\end{tabular*}
\end{table}

\section{Discussion}\label{sec:disc}

The MFA model has been extended to SDB skew-$t$ mixtures. Model development and parameter estimation have been outlined. Two illustrations were presented, using real data, and show that the MSDBFA model can outperform the MSTFA and MFA models. In one case, i.e., the seeds data, the number of components $G$ was fixed to the true number of classes; however, $G$ was selected using the BIC in the other, i.e., the wine data. Of course, the fact that the MSDBFA model outperformed the MSTFA and MFA models herein does not mean that it is a better model in general. Rather, we have shown that the MSDBFA model can outperform these models and so it is a worthy addition to the model-based clustering ``toolkit".

More should be said about the computational challenges to be addressed in fitting the MSDBFA model. Namely, the EM algorithm requires computing several $p$-dimensional integrals on each iteration of the algorithm. As $p$ becomes large, this task becomes increasingly burdensome. Currently, we have implemented code for model fitting using the {\sf R} software. However, developing analogous {\tt C} code, exploring different techniques for evaluating these integrals, and using parallel computing are possible solutions to help address this computational challenge. This remains a problem for future work. 

We note that the SDB formulation of the skew-$t$ distribution is a special case of the so-called canonical fundamental skew-$t$ (CFUST) distribution, introduced by \cite{arellano05}. \cite{lee16} use a mixture of CFUST distributions. Future work will focus on extending the MSDBFA model introduced herein to a mixture of CFUST factor analyzers model. We also note the very recent work of \cite{gallaugher17a} on a formulation of the matrix skew-$t$ distribution using a matrix variate analogue of the generalized hyperbolic formulation of the skew-$t$ distribution used by \cite{murray14b,murray14a}. Future work will also consider matrix analogues of the SDB and canonical fundamental formulations of the skew-$t$ distributions, as well as applying restrictions to the component scale matrices in the MSDBFA model to develop a family of models analogous to the PMSTFA family. The latter direction was not pursued herein because of the computational challenges associated with running many MSDBFA models.

\paragraph{Acknowledgements}
{\small
The authors are grateful to two anonymous reviewers for very helpful comments, which have significantly improved the overall quality of this paper. This work was supported by an Ontario Graduate Scholarship (Murray), respective Discovery Grants from the Natural Sciences and Engineering Research Council of Canada (Browne, McNicholas), and the Canada Research Chairs program (McNicholas).
}

{\small
\bibliographystyle{chicago}

\begin{thebibliography}{}

\bibitem[\protect\citeauthoryear{Aitken}{Aitken}{1926}]{aitken26}
Aitken, A.~C. (1926).
\newblock A series formula for the roots of algebraic and transcendental
  equations.
\newblock {\em Proceedings of the Royal Society of Edinburgh\/}~{\em 45\/}(1),
  14--22.

\bibitem[\protect\citeauthoryear{Andrews and McNicholas}{Andrews and
  McNicholas}{2011a}]{andrews11a}
Andrews, J.~L. and P.~D. McNicholas (2011a).
\newblock Extending mixtures of multivariate $t$-factor analyzers.
\newblock {\em Statistics and Computing\/}~{\em 21\/}(3), 361--373.

\bibitem[\protect\citeauthoryear{Andrews and McNicholas}{Andrews and
  McNicholas}{2011b}]{andrews11b}
Andrews, J.~L. and P.~D. McNicholas (2011b).
\newblock Mixtures of modified t-factor analyzers for model-based clustering,
  classification, and discriminant analysis.
\newblock {\em Journal of Statistical Planning and Inference\/}~{\em 141\/}(4),
  1479--1486.

\bibitem[\protect\citeauthoryear{Arellano-Valle and Genton}{Arellano-Valle and
  Genton}{2005}]{arellano05}
Arellano-Valle, R.~B. and M.~G. Genton (2005).
\newblock On fundamental skew distributions.
\newblock {\em Journal of Multivariate Analysis\/}~{\em 96\/}(1), 93--116.

\bibitem[\protect\citeauthoryear{Azzalini, Browne, Genton, and
  McNicholas}{Azzalini et~al.}{2016}]{azzalini14}
Azzalini, A., R.~P. Browne, M.~G. Genton, and P.~D. McNicholas (2016).
\newblock On nomenclature for, and the relative merits of, two formulations of
  skew distributions.
\newblock {\em Statistics and Probability Letters\/}~{\em 110\/}(201--206).

\bibitem[\protect\citeauthoryear{Azzalini and Capitanio}{Azzalini and
  Capitanio}{2014}]{azzalini14b}
Azzalini, A. and A.~Capitanio (2014).
\newblock {\em The Skew-Normal and Related Families}.
\newblock IMS monographs. Cambridge: Cambridge University Press.

\bibitem[\protect\citeauthoryear{Azzalini and Dalla~Valle}{Azzalini and
  Dalla~Valle}{1996}]{azzalini96}
Azzalini, A. and A.~Dalla~Valle (1996).
\newblock The multivariate skew-normal distribution.
\newblock {\em Biometrika\/}~{\em 83\/}(4), 715--726.

\bibitem[\protect\citeauthoryear{Baek and McLachlan}{Baek and
  McLachlan}{2011}]{baek11}
Baek, J. and G.~J. McLachlan (2011).
\newblock Mixtures of common t-factor analyzers for clustering high-dimensional
  microarray data.
\newblock {\em Bioinformatics\/}~{\em 27\/}(9), 1269--1276.

\bibitem[\protect\citeauthoryear{Baek, McLachlan, and Flack}{Baek
  et~al.}{2010}]{baek10}
Baek, J., G.~J. McLachlan, and L.~K. Flack (2010).
\newblock Mixtures of factor analyzers with common factor loadings:
  Applications to the clustering and visualization of high-dimensional data.
\newblock {\em IEEE Transactions on Pattern Analysis and Machine
  Intelligence\/}~{\em 32}, 1298--1309.

\bibitem[\protect\citeauthoryear{B\"{o}hning, Dietz, Schaub, Schlattmann, and
  Lindsay}{B\"{o}hning et~al.}{1994}]{bohning94}
B\"{o}hning, D., E.~Dietz, R.~Schaub, P.~Schlattmann, and B.~Lindsay (1994).
\newblock The distribution of the likelihood ratio for mixtures of densities
  from the one-parameter exponential family.
\newblock {\em Annals of the Institute of Statistical Mathematics\/}~{\em 46},
  373--388.

\bibitem[\protect\citeauthoryear{Bouveyron and Brunet-Saumard}{Bouveyron and
  Brunet-Saumard}{2014}]{bouveyron14}
Bouveyron, C. and C.~Brunet-Saumard (2014).
\newblock {Model-based clustering of high-dimensional data: A review}.
\newblock {\em Computational Statistics and Data Analysis\/}~{\em 71}, 52--78.

\bibitem[\protect\citeauthoryear{Browne and McNicholas}{Browne and
  McNicholas}{2015}]{browne15}
Browne, R.~P. and P.~D. McNicholas (2015).
\newblock A mixture of generalized hyperbolic distributions.
\newblock {\em Canadian Journal of Statistics\/}~{\em 43\/}(2), 176--198.

\bibitem[\protect\citeauthoryear{Charytanowicz, Niewczas, Kulczycki, Kowalski,
  Lukasik, and Zak}{Charytanowicz et~al.}{2010}]{charytanowicz10}
Charytanowicz, M., J.~Niewczas, P.~Kulczycki, P.~A. Kowalski, S.~Lukasik, and
  S.~Zak (2010).
\newblock Complete gradient clustering algorithm for features analysis of x-ray
  images.
\newblock In {\em Information Technologies in Biomedicine}, Volume~69 of {\em
  Advances in Intelligent and Soft Computing}, pp.\  15--24. Springer Berlin
  Heidelberg.

\bibitem[\protect\citeauthoryear{Dempster, Laird, and Rubin}{Dempster
  et~al.}{1977}]{dempster77}
Dempster, A.~P., N.~M. Laird, and D.~B. Rubin (1977).
\newblock Maximum likelihood from incomplete data via the {E}{M} algorithm.
\newblock {\em Journal of the Royal Statistical Society: Series B\/}~{\em
  39\/}(1), 1--38.

\bibitem[\protect\citeauthoryear{Forina, Armanino, Castino, and Ubigli}{Forina
  et~al.}{1986}]{forina86}
Forina, M., C.~Armanino, M.~Castino, and M.~Ubigli (1986).
\newblock Multivariate data analysis as a discriminating method of the origin
  of wines.
\newblock {\em Vitis\/}~{\em 25}, 189--201.

\bibitem[\protect\citeauthoryear{Fraley and Raftery}{Fraley and
  Raftery}{2002}]{fraley02a}
Fraley, C. and A.~E. Raftery (2002).
\newblock Model-based clustering, discriminant analysis, and density
  estimation.
\newblock {\em Journal of the American Statistical Association\/}~{\em
  97\/}(458), 611--631.

\bibitem[\protect\citeauthoryear{Franczak, Browne, and McNicholas}{Franczak
  et~al.}{2014}]{franczak13}
Franczak, B.~C., R.~P. Browne, and P.~D. McNicholas (2014).
\newblock Mixtures of shifted asymmetric {L}aplace distributions.
\newblock {\em IEEE Transactions on Pattern Analysis and Machine
  Intelligence\/}~{\em 36\/}(6), 1149--1157.

\bibitem[\protect\citeauthoryear{Gallaugher and McNicholas}{Gallaugher and
  McNicholas}{2017}]{gallaugher17a}
Gallaugher, M. P.~B. and P.~D. McNicholas (2017).
\newblock A matrix variate skew-t distribution.
\newblock {\em Stat\/}~{\em 6\/}(1).

\bibitem[\protect\citeauthoryear{Ghahramani and Hinton}{Ghahramani and
  Hinton}{1997}]{ghahramani97}
Ghahramani, Z. and G.~E. Hinton (1997).
\newblock The {E}{M} algorithm for factor analyzers.
\newblock Technical Report CRG-TR-96-1, University of Toronto, Toronto.

\bibitem[\protect\citeauthoryear{Hubert and Arabie}{Hubert and
  Arabie}{1985}]{hubert85}
Hubert, L. and P.~Arabie (1985).
\newblock Comparing partitions.
\newblock {\em Journal of Classification\/}~{\em 2}, 193--218.

\bibitem[\protect\citeauthoryear{Hurley}{Hurley}{2004}]{hurley04}
Hurley, C. (2004).
\newblock Clustering visualizations of multivariate data.
\newblock {\em Journal of Computational and Graphical Statistics\/}~{\em
  13\/}(4), 788--806.

\bibitem[\protect\citeauthoryear{Karlis and Santourian}{Karlis and
  Santourian}{2009}]{karlis09}
Karlis, D. and A.~Santourian (2009).
\newblock Model-based clustering with non-elliptically contoured distributions.
\newblock {\em Statistics and Computing\/}~{\em 19\/}(1), 73--83.

\bibitem[\protect\citeauthoryear{Lawley and Maxwell}{Lawley and
  Maxwell}{1971}]{lawley71}
Lawley and Maxwell (1971).
\newblock {\em Factor Analysis as a Statistical Method\/} (2nd ed.).
\newblock London: Butterworths.

\bibitem[\protect\citeauthoryear{Lee and McLachlan}{Lee and
  McLachlan}{2014}]{lee14}
Lee, S. and G.~J. McLachlan (2014).
\newblock Finite mixtures of multivariate skew $t$-distributions: some recent
  and new results.
\newblock {\em Statistics and Computing\/}~{\em 24\/}(2), 181--202.

\bibitem[\protect\citeauthoryear{Lee and McLachlan}{Lee and
  McLachlan}{2016}]{lee16}
Lee, S.~X. and G.~J. McLachlan (2016).
\newblock Finite mixtures of canonical fundamental skew t-distributions: the
  unification of the restricted and unrestricted skew t-mixture models.
\newblock {\em Statistics and Computing\/}~{\em 26}, 573--589.

\bibitem[\protect\citeauthoryear{Lin}{Lin}{2010}]{lin10}
Lin, T.-I. (2010).
\newblock Robust mixture modeling using multivariate skew-$t$ distributions.
\newblock {\em Statistics and Computing\/}~{\em 20\/}(3), 343--356.

\bibitem[\protect\citeauthoryear{Lin, Ho, and Lee}{Lin et~al.}{2014}]{lin14}
Lin, T.-I., H.~J. Ho, and C.-R. Lee (2014).
\newblock Flexible mixture modelling using the multivariate skew-t-normal
  distribution.
\newblock {\em Statistics and Computing\/}~{\em 24\/}(4), 531--546.

\bibitem[\protect\citeauthoryear{Lin, Lee, and Hsieh}{Lin et~al.}{2007}]{lin07}
Lin, T.~I., J.~C. Lee, and W.~J. Hsieh (2007).
\newblock Robust mixture modeling using the skew-$t$ distribution.
\newblock {\em Statistics and Computing\/}~{\em 17\/}(2), 81--92.

\bibitem[\protect\citeauthoryear{Lin, McLachlan, and Lee}{Lin
  et~al.}{2016}]{lin16}
Lin, T.-I., G.~J. McLachlan, and S.~X. Lee (2016).
\newblock Extending mixtures of factor models using the restricted multivariate
  skew-normal distribution.
\newblock {\em Journal of Multivariate Analysis\/}~{\em 143}, 398--413.

\bibitem[\protect\citeauthoryear{Lin, Wu, McLachlan, and Lee}{Lin
  et~al.}{2015}]{lin15}
Lin, T.-I., P.~H. Wu, G.~J. McLachlan, and S.~X. Lee (2015).
\newblock A robust factor analysis model using the restricted skew-t
  distribution.
\newblock {\em TEST\/}~{\em 24\/}(3), 510--531.

\bibitem[\protect\citeauthoryear{Lindsay}{Lindsay}{1995}]{lindsay95}
Lindsay, B.~G. (1995).
\newblock Mixture models: Theory, geometry and applications.
\newblock In {\em NSF-CBMS Regional Conference Series in Probability and
  Statistics}, Volume~5, pp.\  63--65. California: Institute of Mathematical
  Statistics: Hayward.

\bibitem[\protect\citeauthoryear{McLachlan, Bean, and Jones}{McLachlan
  et~al.}{2007}]{mclachlan07}
McLachlan, G.~J., R.~W. Bean, and L.~B.-T. Jones (2007).
\newblock Extension of the mixture of factor analyzers model to incorporate the
  multivariate $t$-distribution.
\newblock {\em Computational Statistics and Data Analysis\/}~{\em 51\/}(11),
  5327--5338.

\bibitem[\protect\citeauthoryear{McLachlan and Krishnan}{McLachlan and
  Krishnan}{2008}]{mclachlan08}
McLachlan, G.~J. and T.~Krishnan (2008).
\newblock {\em The {EM} Algorithm and Extensions\/} (2nd ed.).
\newblock New York: Wiley.

\bibitem[\protect\citeauthoryear{McLachlan and Lee}{McLachlan and
  Lee}{2016}]{mclachlan16}
McLachlan, G.~J. and S.~X. Lee (2016).
\newblock Comment on ``{O}n nomenclature, and the relative merits of two
  formulations of skew distributions" by {A. Azzalini, R. Browne, M. Genton,
  and P. McNicholas}.
\newblock {\em Statistics \& Probability Letters\/}~{\em 116}, 1--5.

\bibitem[\protect\citeauthoryear{McLachlan and Peel}{McLachlan and
  Peel}{2000}]{mclachlan00a}
McLachlan, G.~J. and D.~Peel (2000).
\newblock Mixtures of factor analyzers.
\newblock In {\em Seventeeth International Conference on Machine Learning}, San
  Francisco, pp.\  599--606.

\bibitem[\protect\citeauthoryear{McNicholas}{McNicholas}{2016a}]{mcnicholas16a}
McNicholas, P.~D. (2016a).
\newblock {\em Mixture Model-Based Classification}.
\newblock Boca Raton: Chapman and Hall/CRC Press.

\bibitem[\protect\citeauthoryear{McNicholas}{McNicholas}{2016b}]{mcnicholas16b}
McNicholas, P.~D. (2016b).
\newblock Model-based clustering.
\newblock {\em Journal of Classification\/}~{\em 33\/}(3), 331--373.

\bibitem[\protect\citeauthoryear{McNicholas, ElSherbiny, McDaid, and
  Murphy}{McNicholas et~al.}{2015}]{mcnicholas15}
McNicholas, P.~D., A.~ElSherbiny, A.~F. McDaid, and T.~B. Murphy (2015).
\newblock {\em pgmm: Parsimonious Gaussian Mixture Models}.
\newblock R package version 1.2.

\bibitem[\protect\citeauthoryear{McNicholas and Murphy}{McNicholas and
  Murphy}{2008}]{mcnicholas08}
McNicholas, P.~D. and T.~B. Murphy (2008).
\newblock Parsimonious {G}aussian mixture models.
\newblock {\em Statistics and Computing\/}~{\em 18}, 285--296.

\bibitem[\protect\citeauthoryear{McNicholas and Murphy}{McNicholas and
  Murphy}{2010}]{mcnicholas10d}
McNicholas, P.~D. and T.~B. Murphy (2010).
\newblock Model-based clustering of microarray expression data via latent
  {G}aussian mixture models.
\newblock {\em Bioinformatics\/}~{\em 26\/}(21), 2705--2712.

\bibitem[\protect\citeauthoryear{McNicholas, Murphy, McDaid, and
  Frost}{McNicholas et~al.}{2010}]{mcnicholas10a}
McNicholas, P.~D., T.~B. Murphy, A.~F. McDaid, and D.~Frost (2010).
\newblock Serial and parallel implementations of model-based clustering via
  parsimonious {G}aussian mixture models.
\newblock {\em Computational Statistics and Data Analysis\/}~{\em 54\/}(3),
  711--723.

\bibitem[\protect\citeauthoryear{McNicholas, McNicholas, and Browne}{McNicholas
  et~al.}{2017}]{smcnicholas17}
McNicholas, S.~M., P.~D. McNicholas, and R.~P. Browne (2017).
\newblock A mixture of variance-gamma factor analyzers.
\newblock In S.~E. Ahmed (Ed.), {\em Big and Complex Data Analysis:
  Methodologies and Applications}, pp.\  369--385. Cham: Springer International
  Publishing.

\bibitem[\protect\citeauthoryear{Meng and van Dyk}{Meng and van
  Dyk}{1997}]{meng97}
Meng, X.-L. and D.~A. van Dyk (1997).
\newblock The {E}{M} algorithm- an old folk song sung to a fast new tune (with
  discussion).
\newblock {\em Journal of the Royal Statistical Society: Series~B\/}~{\em 59},
  511--567.

\bibitem[\protect\citeauthoryear{Montanari and Viroli}{Montanari and
  Viroli}{2010a}]{montanari10a}
Montanari, A. and C.~Viroli (2010a).
\newblock Heteroscedastic factor mixture analysis.
\newblock {\em Statistical Modelling\/}~{\em 10\/}(4), 441--460.

\bibitem[\protect\citeauthoryear{Montanari and Viroli}{Montanari and
  Viroli}{2010b}]{montanari10}
Montanari, A. and C.~Viroli (2010b).
\newblock A skew-normal factor model for the analysis of student satisfaction
  towards university courses.
\newblock {\em Journal of Applied Statistics\/}~{\em 43}, 473--487.

\bibitem[\protect\citeauthoryear{Morris and McNicholas}{Morris and
  McNicholas}{2016}]{morris15}
Morris, K. and P.~D. McNicholas (2016).
\newblock Clustering, classification, discriminant analysis, and dimension
  reduction via generalized hyperbolic mixtures.
\newblock {\em Computational Statistics and Data Analysis\/}~{\em 97},
  133--150.

\bibitem[\protect\citeauthoryear{Murray, Browne, and McNicholas}{Murray
  et~al.}{2014a}]{murray14b}
Murray, P.~M., R.~P. Browne, and P.~D. McNicholas (2014a).
\newblock Mixtures of skew-t factor analyzers.
\newblock {\em Computational Statistics and Data Analysis\/}~{\em 77},
  326--335.

\bibitem[\protect\citeauthoryear{Murray, McNicholas, and Browne}{Murray
  et~al.}{2014b}]{murray14a}
Murray, P.~M., P.~D. McNicholas, and R.~B. Browne (2014b).
\newblock A mixture of common skew-t factor analyzers.
\newblock {\em Stat\/}~{\em 3\/}(1), 68--82.

\bibitem[\protect\citeauthoryear{O'Hagan, Murphy, Gormley, McNicholas, and
  Karlis}{O'Hagan et~al.}{2016}]{ohagan16}
O'Hagan, A., T.~B. Murphy, I.~C. Gormley, P.~D. McNicholas, and D.~Karlis
  (2016).
\newblock Clustering with the multivariate normal inverse {G}aussian
  distribution.
\newblock {\em Computational Statistics and Data Analysis\/}~{\em 93}, 18--30.

\bibitem[\protect\citeauthoryear{{R Core Team}}{{R Core Team}}{2016}]{R16}
{R Core Team} (2016).
\newblock {\em R: A Language and Environment for Statistical Computing}.
\newblock Vienna, Austria: R Foundation for Statistical Computing.

\bibitem[\protect\citeauthoryear{Sahu, Dey, and Branco}{Sahu
  et~al.}{2003}]{sahu03}
Sahu, S.~K., D.~K. Dey, and M.~Branco (2003).
\newblock A new class of multivariate skew distributions with application to
  {B}ayesian regression models.
\newblock {\em Canadian Journal of Statistics\/}~{\em 31}, 129--150.

\bibitem[\protect\citeauthoryear{Schwarz}{Schwarz}{1978}]{schwarz78}
Schwarz, G. (1978).
\newblock Estimating the dimension of a model.
\newblock {\em Annals of Statistics\/}~{\em 6}, 461--464.

\bibitem[\protect\citeauthoryear{Spearman}{Spearman}{1904}]{spearman04}
Spearman, C. (1904).
\newblock The proof and measurement of association between two things.
\newblock {\em The American Journal of Psychology\/}~{\em 15\/}(1), 72--101.

\bibitem[\protect\citeauthoryear{Steane, McNicholas, and Yada}{Steane
  et~al.}{2011}]{steane11}
Steane, M.~A., P.~D. McNicholas, and R.~Y. Yada (2011).
\newblock Model-based classification via mixtures of multivariate t-factor
  analyzers.
\newblock {\em Communications in Statistics -- Simulation and
  Computation\/}~{\em 41\/}(4), 510--523.

\bibitem[\protect\citeauthoryear{Subedi and McNicholas}{Subedi and
  McNicholas}{2014}]{subedi14}
Subedi, S. and P.~D. McNicholas (2014).
\newblock Variational {B}ayes approximations for clustering via mixtures of
  normal inverse {G}aussian distributions.
\newblock {\em Advances in Data Analysis and Classification\/}~{\em 8\/}(2),
  167--193.

\bibitem[\protect\citeauthoryear{Tortora, Franczak, Browne, and
  McNicholas}{Tortora et~al.}{2014}]{tortora14}
Tortora, C., B.~C. Franczak, R.~P. Browne, and P.~D. McNicholas (2014, March).
\newblock Mixtures of multiple scaled generalized hyperbolic distributions.
\newblock {arXiv} preprint {arXiv}:1403.2332v2.

\bibitem[\protect\citeauthoryear{Tortora, McNicholas, and Browne}{Tortora
  et~al.}{2016}]{tortora15}
Tortora, C., P.~D. McNicholas, and R.~P. Browne (2016).
\newblock A mixture of generalized hyperbolic factor analyzers.
\newblock {\em Advances in Data Analysis and Classification\/}~{\em 10\/}(4),
  423--440.

\bibitem[\protect\citeauthoryear{Vrbik and McNicholas}{Vrbik and
  McNicholas}{2012}]{vrbik12}
Vrbik, I. and P.~D. McNicholas (2012).
\newblock Analytic calculations for the {EM} algorithm for multivariate
  skew-mixture models.
\newblock {\em Statistics and Probability Letters\/}~{\em 82\/}(6), 1169--1174.

\bibitem[\protect\citeauthoryear{Vrbik and McNicholas}{Vrbik and
  McNicholas}{2014}]{vrbik14}
Vrbik, I. and P.~D. McNicholas (2014).
\newblock Parsimonious skew mixture models for model-based clustering and
  classification.
\newblock {\em Computational Statistics and Data Analysis\/}~{\em 71},
  196--210.

\bibitem[\protect\citeauthoryear{Wolfe}{Wolfe}{1965}]{wolfe65}
Wolfe, J.~H. (1965).
\newblock A computer program for the maximum likelihood analysis of types.
\newblock Technical Bulletin 65-15, U.S.\ Naval Personnel Research Activity.

\bibitem[\protect\citeauthoryear{Wraith and Forbes}{Wraith and
  Forbes}{2015}]{wraith15}
Wraith, D. and F.~Forbes (2015).
\newblock Location and scale mixtures of {G}aussians with flexible tail
  behaviour: Properties, inference and application to multivariate clustering.
\newblock {\em Computational Statistics and Data Analysis\/}~{\em 90}, 61--73.

\end{thebibliography}

}
\end{document}